\documentstyle[aps,prl,twocolumn,epsf]{revtex}
\begin{document}
\draft
\flushbottom
\twocolumn[
\hsize\textwidth\columnwidth\hsize\csname @twocolumnfalse\endcsname

\title{$T_c$ suppression in co-doped striped cuprates}
\author{C.~Morais Smith $^{1,2}$, A.~H.~Castro Neto$^{3,4}$, and A.~V.~Balatsky$^5$}
\address{$^1$ Institut de Physique Th\'eorique, P\'erolles, CH-1700 Fribourg, Switzerland.\\
$^2$ I Institut f{\"u}r Theoretische Physik, Universit{\"a}t Hamburg, D-20355 Hamburg, Germany. \\
$^3$ Department of Physics, Boston University, 590 Commonwealth Ave, Boston, MA, 02215 \\
$^4$ Department of Physics,University of California, Riverside, CA, 92521 \\
$^5$ Los Alamos National Laboratory, Los Alamos, NM 87545}

\date{\today}
\maketitle
\tightenlines
\widetext
\advance\leftskip by 57pt
\advance\rightskip by 57pt

\begin{abstract}
We propose a model that explains the reduction of $T_c$ due to
the pinning of stripes by planar impurity co-doping in cuprates.
A geometrical argument about the planar fraction of
carriers affected by stripe pinning leads to a
a linear $T_c$ suppression as a function of impurity concentration $z$.
The critical value $z_c$ for the vanishing of superconductivity
is shown to scale like $T_c^2$ in the under-doped regime and becomes
universal in the optimally- and over-doped regimes. Our theory agrees
very well with the experimental data in single- and bi-layer cuprates
co-doped with Zn, Li, Co, etc...
\end{abstract}

]
\narrowtext

One of the most striking properties of high temperature superconductors
(HTSC) is the sensitivity of the critical temperature, $T_c$,
to planar impurities which are
introduced with the substitution of the Cu atoms in the CuO$_2$ planes.
$T_c$ is suppressed with a few percent
of doping almost independently of the magnetic nature of the impurity
\cite{xiao}. In particular it has been shown experimentally
that the HTSC undergo a superconductor to insulator transition due to Zn doping
\cite{fuku}. The fast depression of $T_c$ with Zn has been studied
extensively by NMR, $\mu$SR \cite{msr,nachumi}, and infrared techniques
\cite{basov}
and a heated debate has been developed on the interpretation of
the experimental data \cite{debate}. It has also been established
that this suppression is more robust in the under-doped compounds
\cite{will}.
The reduction of $T_c$ has been assigned to
formation of magnetic defects \cite{maha,ishida}, electron scattering
by disorder in the presence of a d-wave
order parameter \cite{dwave,momo}, unitary scattering \cite{unitary},
local variation of the superconducting gap\cite{cheese},
and weak localization \cite{kim}. However, although the existing theories
describe well the behavior of these systems at low impurity concentration,
they systematically deviate from the experimental data at higher doping
concentration.

In this work we introduce a scenario for the destruction
of superconductivity which is based on the stripe picture. We propose
that Zn pins stripes and slows down their dynamics in a region of size $R$
in its surrounding.
This effect resembles the ``swiss cheese''-like model with Zn sites creating
voids in the superconducting mesh and destroying superfluid density \cite{nachumi}.
In our model, however, the superfluid density remains unaltered upon Zn co-doping,
whereas the stripe inertia increases locally. Since the
stripe mass density $\rho$ is inversely proportional to the hopping energy
$t$ \cite{cris}, a local decrease of $t$ by pinning will enhance
the stripe effective mass, $\rho$. Magnetic studies of the effect of Zn doping
in cuprates indicate that
co-doping indeed leads to charge carrier localization \cite{fuku,bernhard,julien}.
$T_c(x,z)$ is a function of the carrier density $x$ and $z$
and it is determined by the ratio between the 3D superfluid density
$n_s$ and the $ab$-plane charge carriers effective mass $m^*_{ab}$.
Within the stripe picture we must have $T_c \propto n_s / \rho$. Hence,
the pinning of stripes in the vicinity of the impurity implies
a reduction of $T_c$ by the enhancement of $\rho$.
As a result of our model we obtain $T_c(x,z)/T_c(x,0) = 1 - z/z_c(x)$,
where $z_c(x)$ is the critical concentration required to completely suppress superconductivity.
In under-doped compounds, $z_c(x) \propto x^2$ and scales with $T_c^2(x,0)$ in contrast with
the case of ordinary superconductors.
In optimally- and over-doped materials $z_c(x)$ is independent of $x$ and
exhibits universal behavior. Our results give a very good description of
the experimental data in La$_{2-x}$Sr$_x$CuO$_{4}$ (LSCO),
YBa$_2$Cu$_3$O$_{7-\epsilon}$ (YBCO),
and Bi$_2$Sr$_2$Ca$_{1-y}$Y$_y$Cu$_2$O$_{8+\epsilon}$ (BSCO)
co-doped with different impurities, such as Zn, Li, Co, etc.

\begin{figure}
\epsfysize5cm
\hspace{1cm}
\epsfbox{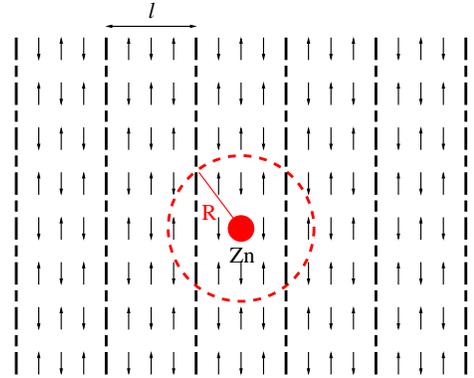}
\vspace{0.5cm}
\caption{Effect of Zn on the stripe grid. Dark dashed lines mark
the regions of stripes with finite superfluid density.
The circle shows the region where hopping is suppressed and the
superfluid effective mass is increased.}
\label{lat}
\end{figure}

Striped phases of holes have been observed experimentally in
the superconductor La$_{1.6-x}$Nd$_{0.4}$Sr$_x$CuO$_{4}$
\cite{tranquada} and in
La$_{2-x}$Sr$_x$NiO$_{4+y}$ \cite{nickel}, which is an antiferromagnetic
insulator. Moreover, stripe formation provides a simple explanation
for the observed magnetic incommensurability in
LSCO \cite{LSCO} and YBCO  \cite{YBCO}.
The magnetic incommensurability appears in neutron scattering by the splitting
of the commensurate peak at ${\bf Q}=(\pi/a,\pi/a)$ ($a \approx 3.8 \AA$
is the lattice spacing) by a quantity $\delta$, which is inversely
proportional to the average inter-stripe distance $\ell$ \cite{yamada}.
In under-doped LSCO, the charge stripes behave as an incompressible
quantum fluid and $\delta$ is proportional to $x$ \cite{zeit}.
In this regime $T_c$ scales with $\delta$, \cite{yamada}
\begin{eqnarray}
T_c(x) = \frac{569}{\ell(x)} \propto \delta
\label{tc}
\end{eqnarray}
where $\ell$ is given in $\AA$.
On the other hand, in the optimally
and over-doped compounds the incommensurability $\delta$
saturates to a constant value and $\ell$ becomes nearly independent
of $x$ \cite{yamada}.
To date, there is strong evidence for stripe formation in
LSCO compounds \cite{yamada,tranquada}.
Recent ion-channeling \cite{venky}, NMR \cite{berthier}, $\mu$SR \cite{akoshima}
and neutron scattering data \cite{mook} also indicate that stripes might be present
in YBCO, whereas a lack of experimental
data still persists for BSCO.
In what follows we will concentrate on the physics of HTSC co-doped with
planar impurities, like La$_{2-x}$Sr$_x$Cu$_{1-z}$Zn$_z$O$_{4}$.
We will verify that our results can describe equally well
YBCO and BSCO co-doped with Zn, Li, and Co.

For simplicity, we assume
that the Zn atoms are located half distance between superconducting stripes and
suppress stripe fluctuations within a circle of radius $R$ around their position,
as shown in Fig.\ \ref{lat}.
This distance $R$ is assumed to scale with the average inter-stripe distance $\ell$. Thus,
we parameterize $R = \gamma \ell/2$ with the doping independent phenomenological parameter
$\gamma > 1$.
As it is shown in Fig.\ \ref{lat}, the stripe length which is pinned per
Zn atom is $4 \sqrt{R^2-\ell^2/4} = 2 \ell \sqrt{\gamma^2-1}$. If we
assume that all the $N_{Zn}$ Zn atoms take part in pinning
the stripes, the total pinned length is $N_{Zn} 2 \ell \sqrt{\gamma^2-1}$.
The transverse kinetic energy density of the stripe is $t/a$, where $t$ is the
single hole kinetic energy.  Thus, the suppressed
energy density in the plane is $\delta \tau = (t/a) 2 \ell \sqrt{\gamma^2-1}N_{Zn} /L^2$
where $L$ is the sample size.
Defining the impurity fraction $z=N_{Zn} (a/L)^2$ and the planar kinetic energy
density $\tau=t/(a \ell)$, we find that
the suppressed energy density reads
\begin{eqnarray}
\delta \tau = \frac{2 z \ell t \sqrt{\gamma^2-1}}{a^3} =
\frac{2 z \ell^2 \sqrt{\gamma^2-1}}{a^2} \tau(x,0).
\label{cluster}
\end{eqnarray}
The energy density of the co-doped system is
given by $\tau(x,z) = \tau(x,0) - \delta \tau$,
which leads to
\begin{eqnarray}
\frac{\tau(x,z)}{\tau(x,0)} =  1 - \frac{z}{z_c(x)}  \, ;
\label{final1}
\end{eqnarray}
where
\begin{eqnarray}
z_c(x) = \frac{1}{2 \sqrt{\gamma^2-1}} \left(\frac{a}{\ell(x)}\right)^2 \, .
\label{zc}
\end{eqnarray}

Superconducting long range order, and therefore $T_c$, is obtained when
the stripe array attains phase coherence. This coherence can be achieved by
the Josephson coupling between stripes which may occur via exchange of
Cooper pairs \cite{kivelson} or d$_{x^2-y^2}$ bosons \cite{antonio}. In this case
the finite transition temperature is in the 2D XY universality class (stabilized
by the inter-planar coupling). Thus, $T_c$ is given by
\begin{equation}
T_c = \frac{\pi \hbar^2 L_c}{2 k_B} \, \frac{n_s}{m^*_{ab}}
\end{equation}
where $L_c$ is the inter-planar distance. Notice that
$n_s \sim 1/(2 a \ell(x) L_c)$ (Ref.\ \cite{tranquada}) and $m^*_{ab}$
is proportional to the stripe linear mass density $\rho = \hbar^2 /(ta^3)$
\cite{cris}. Therefore we conclude that
\begin{equation}
T_c(x,z) \propto \frac{t(z)}{\ell (x)} \propto \tau(x,z).
\label{tc1}
\end{equation}
Combining (\ref{tc1}) with (\ref{final1}) we find:
\begin{eqnarray}
\frac{T_c(x,z)}{T_c(x,0)} = \frac{\tau(x,z)}{\tau(x,0)} =
1 - \frac{z}{z_c(x)}.
\label{tczzc}
\end{eqnarray}

\begin{figure}
\epsfxsize=8cm
\hspace{1cm}
\epsfbox{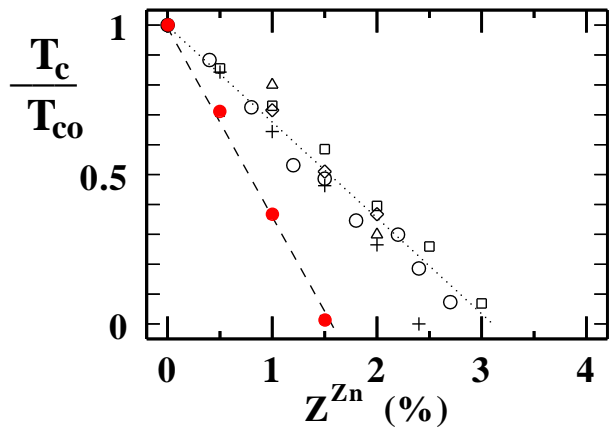}
\vspace{0.5 cm}
\caption{$T_c (x,z)/T_c(x,0)$ versus the Zn doping concentration $z$
for the lanthanate-compound. Dashed and dotted lines correspond to Eq.\
(\ref{tczzc}) with $z_c$ given by Eqs.\ (\ref{zc1}) and (\ref{zc2}),
respectively. Red circles denote x= 0.10 \protect\cite{hara} and the opened
symbols are:
+ x = 0.15 \protect\cite{hara}, circles x = 0.15 \protect\cite{xiao},
lozenges x = 0.18 \protect\cite{yoshi}, triangles x = 0.20 \protect\cite{yoshi},
squares x = 0.20 \protect\cite{hara}.
}
\label{lineartc}
\end{figure}

It is very illuminating to compare the expressions for $z_c(x)$, that is,
the critical Zn doping for which $T_c$ vanishes, in the various regimes.
For under-doped lanthanates ($x<1/8$),
neutron scattering data show that $T_c(x,0) \propto \delta$ \cite{yamada}.
Comparing (\ref{zc}) with (\ref{tc}), one finds an unforeseen result,
which is
\begin{eqnarray}
z_c(x) = \left(\frac{a(\AA)}{805}\right)^2
\frac{T_c^2(x,0)}{\sqrt{\gamma^2-1}}
\label{zc1}
\end{eqnarray}
in the under-doped regime. Moreover, in the optimally and over-doped compounds
the average stripe separation saturates to a value $\ell_s$, and therefore we
expect $z_c$ to be an universal constant, independent of $x$,
\begin{eqnarray}
z_c = \frac{1}{2 \sqrt{\gamma^2-1}} \left(\frac{a}{\ell_s}\right)^2.
\label{zc2}
\end{eqnarray}

\begin{figure}
\epsfxsize=8cm
\hspace{1cm}
\vspace{0.5 cm}
\epsffile{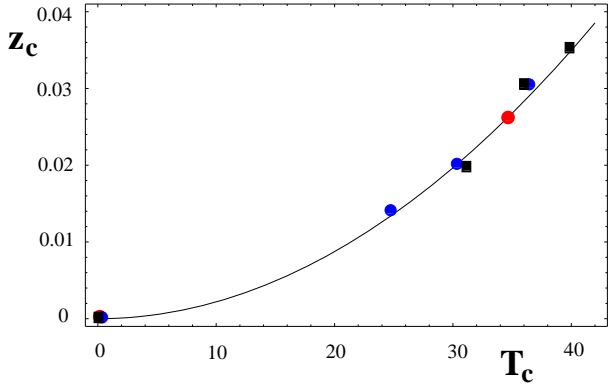}
\caption{$z_c \times T_c$: the continuous line is given by Eq.\ (\ref{zc1}).
The data points are experimental values for under-doped La-compound:
black squares (Ref.\ \protect\cite{momo}); blue circles (Ref.\
\protect\cite{kluge}), red circle (Ref.\ \protect\cite{xiao}).}
\label{zcxtc2}
\end{figure}

In Fig.\ \ref{lineartc}, we plot
the experimental data for LSCO co-doped with Zn \cite{kluge}.
The dashed and dotted lines are the theoretical result (\ref{tczzc})
with $z_c$ given by Eqs.\ (\ref{zc1}) and (\ref{zc2}), respectively.
The agreement between theory and experiment is excellent. One can
easily observe that indeed the data for different $x$ collapse into a
single line for optimally and over-doped compounds. In addition,
in Fig.\ \ref{zcxtc2} we show the experimental data for $z_c \times T_c$, as
well as the theoretical prediction (\ref{zc1}). In both figures, we
used $\gamma \sim 1.42$ so that $R \sim 0.71 \ell $.
Hence, for the optimally- and over-doped materials, for which
the mean stripe distance $\ell_s = 4 a = 15.2$\AA, we estimate
$R \sim 10$\AA. The value of $R$ can be independently
obtained in neutron scattering experiments in Zn doped samples. At this
point in time, however, we are not aware of any data in this regard for LSCO.

\begin{figure}
\epsfysize+13cm
\epsffile{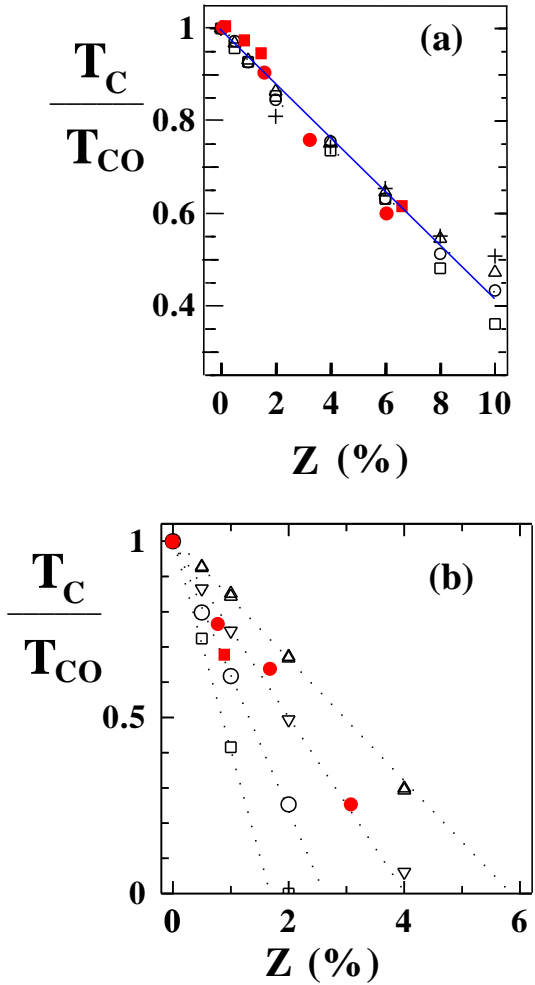}
\vspace{0.5 cm}
\caption{ NMR and $\mu$SR data for ytrium-and bismuth-based compounds. Red
points correspond to YBa$_2$Cu$_3$O$_{7-x}$ \protect\cite{bobroff} co-doped with
Li (squares) and Zn (circles), whereas open symbols correspond to
Bi$_2$Sr$_2$Ca$_{1-y}$Y$_{y}$(Cu$_{1-z}$Co$_z$)$_2$O$_{8+x}$
\protect\cite{kluge}. (a) optimally- and over-doped compounds;
(b) under-doped materials.}
\label{YBi}
\end{figure}

The excellent agreement between theory and experiment is however not
restricted to the Zn doped lanthanates. In Fig.\ \ref{YBi} we display
the experimental data for YBCO (red points)
co-doped with Li (squares) and Zn (circles) \cite{bobroff},
as well as for BSCO co-doped with Co (black points)
\cite{kluge}. The universal behavior at high doping is very clear
(Fig.\ \ref{YBi}a) as well as the different slopes characteristic of the
under-doped compounds (Fig.\ \ref{YBi}b). Observe that YBCO and BSCO
exhibit exactly the same features in terms of Zn doping, with the same value
of $z_c$ in the over-doped regime. We assign this behavior to the fact
that both systems are bi-layers while LSCO is a single-layer system.
Moreover, it has been shown recently that the relationship of $T_c$ and
$\delta$ for YBCO is similar to the one in LSCO \cite{marai}.
Using the value of the incommensurability $\delta = 0.11$ measured by
neutron scattering in YBCO$_{6.6}$ \cite{mook}, which has $T_c = 60$K, we
can determine the stripe separation in this under-doped compound,
$\ell (x) \sim 4.55 a$. Then, by fitting the data from Ref. \cite{bobroff}
for the same compound (red circles in Fig.\ 4b, $z_c = 0.04$) to Eqs.\
(\ref{tczzc}) and (\ref{zc}), we determine the only free parameter in our
model, $\gamma = 1.17$, corresponding to $R \sim 10$ \AA. Now, we use this
value of $\gamma$ to analyze the over-doped regime. The universal line
in Fig.\ 4a defines the critical doping in over-doped bi-layer cuprates,
$z_c \sim 0.13$. By replacing $z_c$ and $\gamma$ in Eq.\ (\ref{zc2}), we
can estimate the saturation value of the average stripe distance. We then
find $\ell_s \sim 3a$, corresponding to $R = \gamma \ell_s / 2 \sim 6$ \AA.
These results agree with magnetic measurements in over-doped YBCO with Zn
impurities: NMR studies
suggest the existence of two different relaxation times ascribed to
$^{63}$Cu sites away from and near Zn impurities \cite{ishida} and
inelastic neutron scattering data \cite{sidis} indicate that Zn induces a
magnetic perturbation on a range of $R \sim 7 \AA$, very close to our
estimates. Moreover, we
predict that the saturation of the incommensurability in YBCO
should start at optimal doping, where $\delta_s \sim 0.2$ and $\ell_s \sim 3a$,
in contrast with the case of LSCO where the saturation starts at
$x \sim 0.12$ and $\ell_s \sim 4a$.
This result agrees with recent neutron scattering data
in YBCO \cite{marai} and establishes a very important relationship between
two different kinds of experiments, namely, magnetization and neutron
scattering. By determining the doping concentration $x$ beyond which the
$T_c(x,z)/T_c(x,0)$ versus $z$ line becomes universal, one is automatically
finding the value of $x$ beyond which a saturation should be reached in the
$\delta$ versus $x$ plot.
It is also a straightforward conclusion of the data in Fig.\ \ref{YBi} that
BSCO has a behavior very similar to YBCO although there is no direct
evidence for incommensurability or stripes in BSCO. It is interesting to
notice that $T_c(x,z)/T_c(x,0)$ when plotted versus the in-plane 2D
residual resistance, $\rho_0^{2D}$, exhibits universal behavior in the
under-doped but not in the over-doped regime \cite{fuku}. However, as a function
of $z$, as we have shown here, the universal behavior is found in the
optimally- and over-doped regimes. This fact is due to the linear dependence
of $\rho_0^{2D}$ with $z$ \cite{fuku}.

We finally would like to point out that the prediction that $z_c \propto T^2_c(x, z=0)$
 is highly unusual. In BCS-like models the Abrikosov-Gor'kov \cite{AG} approach
predicts that $T_c$ is
suppressed when the scattering rate $1/\tau_s \propto z$ becomes of the order
of the superconducting gap $\propto T_c(x,0)$ and therefore one would expect
$z_c \propto T_c(x, z=0)$ in disagreement with the data.

In conclusion, we presented a  ``geometrical'' model for the suppression
of $T_c(x,z)$ by planar impurities within the stripe model.
It is based on the assumption that Zn impurities work as local pinning centers
and suppress the transverse kinetic energy density of the stripes in their
immediate vicinity. Our model leads to a linear $T_c$ suppression with doping, with a
slope inversely proportional to the critical impurity concentration $z_c$,
as given by Eq.\ (\ref{tczzc}). Moreover, we find that $z_c$ exhibits
a non-trivial behavior in under-doped compounds, $z_c(x) \propto T^2_c(x)$, and
a universal one, $z_c = $const. at optimally- and over-doped compositions.
The obtained results agree very well with the experimental findings and allow
to establish a connection between neutron scattering and magnetization measurements.

We are grateful to M. Akoshima and Y. Koike for sending us their experimental data
and to M.\ Vojta, K.\ Yamada, and T.\ Egami for useful discussions. The Aspen Center for
Physics, where this work was initiated, is acknowledged for its hospitality.
A.~H.~C.~N. acknowledges partial support provided by the Collaborative University of
California - Los Alamos (CULAR)  research grant under the auspices of the US Department
of Energy. Work of A.~V.~B. was supported by the US DOE.


\begin{references}

\bibitem{xiao}K.~Westerholt {\it et al}., Phys.~Rev.~B {\bf 39},
11680 (1989)
I.~Felner and B.~Brosh, Phys.~Rev.~B {\bf 43}, 10364 (1991);
Y.~Takano {\it et al.}, Physica B {\bf 169}, 683 (1991);
G.~Xiao {\it et al.},  Phys.~Rev.~B {\bf 42}, 8752 (1990); {\it idem}
J.~Low Temp.~Phys. {\bf 105}, 521 (1996).

\bibitem{fuku}Y~Fukuzumi {\it et al}., Phys.~Rev.~Lett. {\bf 76}, 684 (1996).

\bibitem{msr}G.~V.~M.~Williams {\it et al}., Phys~Rev.~B {\bf 51}, 16503 (1995);
C.~Bernhard {\it et al}., Phys.~Rev.~Lett. {\bf 77}, 2304 (1996).

\bibitem{nachumi} B.~Nachumi {\it et al.}, Phys.~Rev.~Lett. {\bf 77}, 5421
(1996).

\bibitem{basov}D.~N.~Basov {\it et al}.,Phys. Rev Lett. {\bf 81}, 2132 (1998).
\bibitem{debate}C~Bernhard {\it et al}., Phys.~Rev.~Lett. {\bf 80}, 205 (1998);
B.~Nachumi {\it et al.}, Phys.~Rev.~Lett. {\bf 80}, 206 (1998).

\bibitem{will}
J.~L.~Tallon {\it et al}., Phys.~Rev.~Lett. {\bf 79}, 5294 (1997).

\bibitem{maha}
A.~V.~Mahajan {\it et al}., Phys.~Rev.~Lett. {\bf 72}, 3100
(1994);
S.~Zagoulaev {\it et al}., Phys.~Rev.~B {\bf 52}, 10474 (1995).

\bibitem{ishida}
K.~Ishida {\it et al}., J.~Phys.~Soc.~Japan {\bf 62}, 2803 (1993).

\bibitem{dwave}K.~Ishida {\it et al}., Physica C {\bf 179}, 29 (1991);
R.~E.~Walstedt {\it et al.}, Phys.~Rev.~B {\bf 48}, 10646 (1993).

\bibitem{momo}N.~Momono {\it et al}., Physica C {\bf 233}, 395 (1994).

\bibitem{unitary}K.~A.~Mirza {\it et al}., Physica C {\bf 282}, 1405 (1997).

\bibitem{cheese}M.~Franz {\it et al}., Phys.~Rev.~B {\bf 56}, 7882 (1997).

\bibitem{kim} Y.-J.\ Kim and K.\ J.\ Chang, cond-mat/9801071.

\bibitem{cris}  C.\ Morais Smith \textit{et al.}, Phys.\ Rev.\ B {\bf 58},
453 (1998).

\bibitem{bernhard} C.\ Bernhard {\it et al.}, Phys.~Rev.~B {\bf 58}, R8937
(1998).

\bibitem{julien} M.-H.\ Julien {\it et al.}, Phys.~Rev.~Lett. {\bf 84}, 3422
(2000).

\bibitem{tranquada}J.~M.~Tranquada {\it et al.}, Nature, {\bf 375}, 561,
(1995).

\bibitem{nickel}J.~M.~Tranquada {\it et al.}, Phys.~Rev.~Lett. {\bf 73}, 1003
(1994).

\bibitem{LSCO}S.~W.~Cheong {\it et al.}, Phys.~Rev.~Lett. {\bf 67}, 1791
(1991).

\bibitem{YBCO}P.~Dai, H.\ A.\ Mook, and F.\ Dogan, Phys.~Rev.~Lett. {\bf 80},
1738 (1998);  H.\ A.\ Mook {\it et al.}, Nature {\bf 395}, 580 (1998);
{\it idem} {\bf 401}, 145 (1999).

\bibitem{yamada}K.~Yamada {\it et al}., J.~Superconductivity {\bf 10},
343 (1997); B.~O.~Wells {\it et al}., Science {\bf 277}, 1067 (1997).

\bibitem{zeit}A.~H.~Castro Neto, Zeits.~Phys.~B {\bf 103}, 185 (1997).


\bibitem{venky} R.\ P.\ Sharma {\it et al.}, Nature {\bf 404}, 736 (2000).

\bibitem{berthier} B.\ Gr\'evin, Y.\ Berthier, and C.\ Collin, Phys.\ Rev.\ Lett.
{\bf 85}, 1310 (2000).

\bibitem{akoshima} M.\ Akoshima {\it et al.}, J.\ Low Temp.\ Phys.\
{\bf 117}, 1163 (1999); {\it idem},  Phys.~Rev.~B in press (2000).

\bibitem{mook} H.\ A.\ Mook, P.\ Dai, F.\ Dogan, and R.\ D. Hunt,
Nature {\bf 404}, 729 (2000).

\bibitem{kivelson}V.~J.~Emery {\it et al.},
Phys. Rev. Lett. {\bf 85}, 2160 (2000).

\bibitem{antonio}A.~H.~Castro Neto, cond-mat/0007434.

\bibitem{hara} H.\ Harashina {\it et al.},  Physica C {\bf 212}, 142 (1993).

\bibitem{yoshi} R.\ Yoshizaki {\it et al.},  Physica C {\bf 166}, 417 (1990).

\bibitem{kluge}T.~Kluge {\it et al}., Phys.~Rev.~B {\bf 52}, R727 (1995).

\bibitem{bobroff} J.\ Bobroff {\it et al.},Phys.~Rev.~Lett. {\bf 83}, 4381 (1999);
cond-mat/0008396.

\bibitem{marai}A.\ V.\ Balatsky and P.\ Bourges, Phys.~Rev.~Lett. {\bf 82}, 5337 (1999);
M.~Arai {\it et al.}, cond-mat/9912233.

\bibitem{sidis} Y.\ Sidis {\it et al.}, Phys.~Rev.~B {\bf 53}, 6811 (1996).

\bibitem{AG} A.A.Abrikosov and L.P.Gor'kov, Soviet Phys. JETP {\bf 12},1243
(1961).

\end{references}
\end{document}